\documentclass[12pt]{article}
\newcommand{\la}[1]{\label{#1}}

\newcommand{\vecm}{{\vec{\bf m}}}
\newcommand{\vecu}{{\vec{\bf u}}}
\newcommand{\vect}{{\vec{\bf t}}}
\newcommand{\vecv}{{\vec{\bf v}}}

\newcommand{\vecr}{{\vec{\bf r}}}

\newcommand{\bfe}{{\bf e}}

\newcommand{\be}{\begin{equation}}
\newcommand{\ee}{\end{equation}}
\newcommand{\ba}{\begin{eqnarray}}
\newcommand{\ea}{\end{eqnarray}}
\newcommand{\bastar}{\begin{eqnarray*}}
\newcommand{\eastar}{\end{eqnarray*}}

\begin{document}
\begin{titlepage}

\vskip 2.0truecm

\begin{center}
$ ~$
\end{center}


\begin{center}
{ 
\bf \large \bf  ASPECTS OF ELECTRIC AND MAGNETIC
\\ \vskip 0.2cm
VARIABLES IN SU(2) YANG-MILLS THEORY \\
}
\end{center}

\vskip 0.5cm

\begin{center}
{\bf Ludvig Faddeev$^{\, *}$ } {\bf \ and \ } {\bf
Antti J. Niemi$^{\, ** }$ } \\

\vskip 0.6cm

{\it $^*$St.Petersburg Branch of Steklov Mathematical
Institute \\
Russian Academy  of Sciences, Fontanka 27 , St.Petersburg, 
Russia$^{\, \ddagger}$ } \\

\vskip 0.3cm

{\it $^{**}$Department of Theoretical Physics,
Uppsala University \\
P.O. Box 803, S-75108, Uppsala, Sweden } \\

\end{center}

\vskip 4.0cm
\noindent
We introduce a novel decomposition of the four dimensional
SU(2) gauge field. This decomposition realizes explicitely
a symmetry between electric and magnetic variables,
suggesting a duality picture between the corresponding phases.
It also indicates that at large distances the Yang-Mills 
theory involves a three component unit vector field, a 
massive Lorentz vector field, and a neutral scalar field
that condenses which yields the mass scale. Our results are 
consistent with the proposal that the physical spectrum of 
the theory contains confining strings which are 
tied into stable knotted solitons.
\vfill

\begin{flushleft}
\rule{5.1 in}{.007 in} \\
$^{*}$ \hskip 0.2cm {\small  E-mail: \scriptsize
\bf FADDEEV@PDMI.RAS.RU } \\
\hskip 0.5cm supported by RFFR 99-01-00101, INTAS 99-01705, STINT IG2001-062
and the \\
\hskip 0.5cm Swedish Royal Academy of Sciences \\
\vskip 0.2cm
$^{**}$  {\small E-mail: \scriptsize
\bf NIEMI@TEORFYS.UU.SE} \\
\hskip 0.5cm supported by NFR Grant F-AA/FU 06821-308 \\
\end{flushleft}
 
\end{titlepage}

In the infrared limit the phase structure of 
a four dimensional Yang-Mills theory is expected 
to be nontrivial. In particular, there should be
a mass gap and color should be confined  
\cite{thf}. Presumably this also bestows 
new collective variables, more appropriate for 
describing the long distance theory 
than the gauge field $A_\mu^a$ which relates to 
the short distance spectrum. 
In a series of articles \cite{prl}, \cite{ed} 
we have proposed that the infrared limit of SU(2) Yang-Mills 
theory involves a variant of the 3+1 dimensional 
nonlinear $\sigma$-model with a dynamical 
field $\vecr(x)$ being a unit three vector. 
The ensuing Lagrangian should
contain the following three terms, 
\be
L \ = \ (\partial_\mu \vecr )^2
\ - \ \Lambda ( \vecr \cdot \partial_\mu 
\vecr \times \partial_\nu
\vecr )^2 \ - \ V(\vecr) 
\la{fada}
\ee
Here the first term from the left is the standard
nonlinear $\sigma$-model contribution. The second
term was introduced in \cite{fadde}. Together 
with the $\sigma$-model term it yields an energy
functional that supports knotted solitons
\cite{nature,jarmo}. The third term is a 
potential term. With it, the Lagrangian
has the functional form of a 2+1 dimensional 
baby Skyrme model \cite{zakr}. But 
in three spatial dimensions the potential 
term has no effect on the stability of the 
knotted solitons which are described by the first two terms 
in (\ref{fada}). The potential term does influence
the shape of the energy density and the mutual 
interactions of these solitons. But more importantly,
it breaks the global SO(3) symmetry of the two first 
terms which removes the massless Goldstone bosons 
that would be present otherwise. This is necessary
since there are no massless particles in the 
infrared spectrum of the Yang-Mills theory. 

In the present Letter we shall
argue that the Lagrangian (\ref{fada}), in combination
with a massive Lorentz vector and a neutral scalar, may indeed
be relevant in describing the infrared limit of
3+1 dimensional SU(2) Yang-Mills theory. Our starting point 
is a {\it novel} decomposition of the gauge field $A^a_\mu$,
a modification of our previous decomposition \cite{prl}
that extended the earlier one by Cho \cite{cho2}.
The decomposition in \cite{prl,cho2} entails a three 
component unit vector $\vecr$ with a natural magnetic 
interpretation. But the string-like excitations of 
a gauge theory should relate to variables with a natural 
electric interpretation \cite{thf}. For that reason
we now introduce a {\it different} decomposition of 
the gauge field. In fact, the decomposition that we
present here involves two sets of variables which can
be viewed as electric and magnetic respectively.
These variables enter in a very symmetric manner, 
which leads to a duality 
picture between them. In the first part of the Letter
we describe our new decomposition and show
how it leads to structures akin those present in 
(\ref{fada}). We then continue with a 
somewhat more speculative discussion how one
could actually relate the
infrared SU(2) Yang-Mills theory to the Lagrangian
(\ref{fada}).

We shall consider a four dimensional (Euclidean space for the
moment) SU(2) Yang-Mills theory. In the so-called maximal 
abelian gauge which is very popular
in lattice studies, one treats
the Cartan $A_\mu^3$ as a U(1) gauge 
field while $A_\mu^+ = A_\mu^1 + i A_\mu^2$ together
with its complex conjugate are charged vector fields.
The two vector fields $A_\mu^1$ and $A_\mu^2$ lie in
a plane of a four dimensional space, and this plane can
be parametrized by a {\it twobein} ${e^a}_\mu$ ($a=1,2$)
with
\[
{e^a}_\mu {e^b}_\mu \ = \ \delta^{ab}
\]
We can then write the $a=1,2$ components as  
\be
A^a_\mu \ = \ {M^a}_b {e^b}_\mu
\la{dec1}
\ee
But the two off-diagonal components of $A_\mu^a$
describe eight 
field degrees of freedom while on the {\it r.h.s.} of 
(\ref{dec1}) we have nine since the 
matrix ${M^a}_b$ has four independent 
elements and the two normalized vectors 
${e^a}_\mu$ have five independent components. However,
there is also an internal SO(2) $\sim$ U(1) rotation invariance between 
${M^a}_b$ and ${e^a}_\mu$: If for a fixed $\mu$ we rotate 
${e^a}_\mu$ according to
\[
{e^a}_\mu \ \to \ {{\cal O}^a}_b {e^b}_\mu
\]
the decomposition (\ref{dec1}) remains intact provided we 
also transpose ${M^a}_b$ from the right with the same
${{\cal O}^a}_b$. When we account for this gauge invariance,
each side in (\ref{dec1}) indeed involves eight independent
field degrees of freedom.

We introduce the combination
\be
\bfe_\mu \ = \ \frac{1}{\sqrt{2}} ( {e^1}_\mu + i {e^2}_\mu)
\la{cple}
\ee
so that we have 
\be
\bfe^2 = 0 \ \ \ \ \ \ \ \ \& 
\ \ \ \ \ \ \ \ \bfe \cdot \bfe^* = 1
\la{norme}
\ee
We re-write the decomposition (\ref{dec1}) as
\be
A^1_\mu + i A^2_\mu \ = \ i 
\psi_1 \bfe_\mu \ + \ i \psi_2 \bfe_\mu^\star
\la{rhoA}
\ee
where we have arranged the four matrix elements of 
${M^a}_b$ into two complex scalar 
fields $\psi_{1}$ and $\psi_2$.
A diagonal SU(2) gauge transformation sends 
\be
A^3_\mu  \equiv A_\mu \to A_\mu - \partial_\mu \xi
\la{vec1a}
\ee
and multiplies both $\psi_1$ and $\psi_2$ by a common 
phase,
\be
\psi_{1,2} \ \to \ e^{i\xi} \psi_{1,2}
\la{vec1b}
\ee
but leaves ${e^a}_\mu$ intact. This is the natural 
action of a vector-like, or electric U(1) gauge transformation
with $\psi_{1,2}$ the electrically charged fields.
On the other hand, under the internal 
U(1) rotation we have  
\[    
\bfe_\mu \ \to \ e^{-i\zeta} \bfe_\mu 
\]
and 
\[
\matrix{ \psi_1 \ \to \ e^{i\zeta}\psi_1  \cr
\psi_2 \ \to \ e^{-i\zeta}\psi_2 }
\]
Now the decomposition (\ref{rhoA}) remains
intact, while the composite vector field
\be
C_\mu \ = \ i \bfe \cdot \partial_\mu {\bfe}^\star
\la{defC}
\ee
transforms according to
\be
C_\mu \ \to \ C_\mu \ - \ \partial_\mu \zeta
\la{gauC}
\ee
Hence $C_\mu$ can be viewed as a gauge field 
for the internal rotation. In particular
(\ref{gauC}) admits a natural interpretation 
as an axial-like or magnetic U(1) gauge transformation.

We employ the complex vector (\ref{cple}) to define a real
antisymmetric tensor
\be
G_{\mu\nu} \ = \ i (\bfe_\mu \bfe_\nu^\star - 
\bfe_\nu \bfe_\mu^\star )
\la{Gij}
\ee
which is invariant under the electric and magnetic gauge 
transformations. We introduce the corresponding
Maxwellian electric and magnetic combinations
\be
{\cal E}_k \ = \ G_{k0} \ \ \ \ \ \ \& \ \ \ \ \ \
{\cal B}_k \ = \ \frac{1}{2} \epsilon_{klm} G_{lm}
\la{maxeb}
\ee
Then 
\be
\vecu \ = \ \vec{\cal E} \ + \ \vec{\cal B}
\ \ \ \ \ \ \& \ \ \ \ \ \
\vecv \ = \ \vec{\cal E} \ - \ \vec{\cal B}
\la{defuv}
\ee 
are two independent three-component unit vectors. When 
we invert (\ref{maxeb})
to give $\bfe_\mu$ in terms of the vectors $\vecu$
and $\vecv$ we get
\be
\bfe_\mu \ = \ \frac{e^{i\phi}}{\sqrt{2}} \left
( \ e_0 \ , \ \frac{1}{2e_0} [ \vecu \times \vecv + i ( \vecu +
\vecv) ] \ \right) 
\la{euv1}
\ee
Here $\phi$ is the phase of the $\mu = 0$
component
of $\bfe_\mu$ while the normalization condition
(\ref{norme}) yieds for the modulus
\[
e_0 \ = \ \sqrt{ 1 + \vecu \cdot \vecv}
\]
For the magnetic gauge field (\ref{gauC}) this gives
\be
C_\mu \ = \ \frac{1}{1 + \vecu \cdot \vecv} \
(\partial_\mu \vecu \ + \ \partial_\mu \vecv) \cdot
\vecu \times \vecv \ + \ 2\partial_\mu \phi
\la{C2nd}
\ee
which identifies $\phi$ as the magnetic phase. 
We also introduce a pair of complex vectors
\be
U_\mu \ = \ e^{i\phi}
\frac{\partial_\mu \vecu \cdot 
( \vecv + i \vecu \times \vecv ) }{\sqrt{ 
1 - (\vecu \cdot \vecv)^2}}
\la{Q1def}
\ee
\be
V_\mu \ = \ e^{i\phi} \frac{\partial_\mu 
\vecv \cdot ( \vecu + i 
\vecu \times \vecv )}{\sqrt{ 1 
- (\vecu \cdot \vecv)^2}}
\la{Q2def}
\ee
Then 
\be
\partial_\mu \bfe \cdot \partial_\mu \bfe \ = \ 
U_\mu V_\mu
\la{ee}
\ee
Finally, we set $\rho^2 = |\psi|^2$ and define the 
three-component unit vector
\[
\vect \ = \ \frac{1}{\rho^2} \
(\psi_1^\star \ \psi_2^\star ) \ \vec 
\sigma \ \pmatrix{ \psi_1 \cr \psi_2 }
\]
where $\vec \sigma$ are the standard Pauli matrices. 
This vector is invariant under the electric gauge 
transformation. The component $t_3$ 
is also invariant under the magnetic gauge transformation,
but for the other two components we have
\be
t_{\pm} \ = \ \frac{1}{2} ( t_1 \pm i t_2 ) \ 
\to \ e^{\mp 2i \zeta} t_{\pm}
\la{tphase}
\ee

With these definitions we now proceed to the 
Yang-Mills action where we impose
a partial gauge fixing, only for the off-diagonal components.
For this we consider the following gauge fixed 
Lagrangian \cite{thf}
\be
L_{YM} \ = \ \frac{1}{4} (F^a_{\mu\nu})^2 \
+ \ \frac{1}{2}
[ ( \partial_\mu \delta^{ab}
-  \epsilon^{ab}A_\mu ) A^b_\mu]^2
\la{gfym}
\ee
Here we have a renormalizable background gauge
condition for the off-diagonal components $A_\mu^{\pm}$, with
respect to the diagonal Cartan component $A_\mu \equiv A_\mu^3$
of the gauge field. The ensuing Lagrangian for the ghosts
is constructed in an entirely standard fashion. 
But since it only becomes relevant 
in computing radiative corrections,
we do not write the ghost contribution explicitely.

We substitute the decomposed gauge field in (\ref{gfym})
and find for the gauge fixed Lagrangian
\be
L_{YM} \ = \ 
\frac{1}{4} F_{\mu\nu}^2 \ + \
|D_\mu^{ab}\psi_b|^2 \ + \ \frac{1}{8} ( |\psi_1|^2 - 
|\psi_2|^2)^2
\la{fulL1}
\ee
\be
+ \ \frac{1}{2}\rho^2 ( |\partial_\mu \vecu |^2 \ + 
\ |\partial_\mu \vecv |^2) 
\ + \ \frac{1}{2} \rho^2 t_- U_\mu V_\mu
\ + \ \frac{1}{2} \rho^2 t_+ U_\mu^*
V_\mu^* 
\la{fulL2}
\ee
\be
+ \ \frac{1}{2} \rho^2 t_3 F_{\mu\nu} G_{\mu\nu}
\ + \ ghosts
\la{fulL3}
\ee
Here
\be
D_\mu^{ab} \ = \ \delta^{ab} ( \partial_\mu + i A_\mu)
\ - \ i \sigma_3^{ab} C_\mu
\la{covder}
\ee
is the U(1)$\times$U(1) covariant derivative. Indeed, we
note that (\ref{fulL1})-(\ref{fulL3}) is invariant both
under the (electric) U(1) of the SU(2) gauge group,
and under the internal (magnetic) U(1).

We define the vector field
\be
B_\mu \ = \ A_\mu \ + \ \frac{i}{2\rho^2}
\left[ \ \psi_a {\hat D}^{ab}_\mu \bar\psi_b - 
\bar\psi_a {\hat D}^{ab}_\mu \psi_b \ \right] \ \equiv \
A_\mu \ + \ \frac{i}{2\rho^2} J_\mu
\la{shift}
\ee
where
\[
{\hat D}^{ab} \ = \delta^{ab} \partial_\mu
\ - \ i \sigma_3^{ab} C_\mu
\]
is the magnetic covariant derivative, {\it i.e.} 
(\ref{covder}) with $A_\mu$ removed; notice that  
$B_\mu$ is invariant under 
the electric U(1) gauge transformation. With this we
then get for the Lagrangian
(\ref{fulL1})-(\ref{fulL3})
\be
L_{YM} \ = \ 
\frac{1}{4} ( H_{\mu\nu} + M_{\mu\nu} +  
K_{\mu\nu} t_3 )^2 \ + \ \frac{1}{2} (\partial_\mu \rho)^2
\ + \ \rho^2 (\nabla^{ij}_\mu t_j )^2 
\ + \ \rho^2 B_\mu^2
\ + \ \frac{1}{8} \rho^4 t_3^2 
\la{fulL4}
\ee
\be
+ \ \frac{1}{2}\rho^2 ( |\partial_\mu \vecu |^2 \ + 
\ |\partial_\mu \vecv |^2) 
\ + \ \frac{1}{2} \rho^2 ( t_- U_\mu V_\mu
\ + \ t_+ U_\mu^* V_\mu^* \ + \ t_3 [  H_{\mu\nu} + M_{\mu\nu} +  
K_{\mu\nu} t_3 ] G_{\mu\nu} )
\ + \ ghosts
\la{fulL5}
\ee
where
\be
\nabla_\mu^{ij} \ = \ \delta^{ij} \partial_\mu
+ 2 \epsilon^{ij3} C_\mu
\la{covvec}
\ee
describes the action of the magnetic covariant derivative
on a vector, and
\[
H_{\mu\nu} \ = \ \partial_\mu B_\nu - \partial_\nu B_\mu
\]
\be
M_{\mu\nu} \ = \ \epsilon^{ijk} t_i 
\nabla^{jl}_\mu t_l
\nabla^{km}_\nu t_m
\la{K}
\ee
\[
K_{\mu\nu} \ = \ \partial_\mu C_\nu - \partial_\nu C_\mu
\]

The Lagrangian (\ref{fulL4}), \ref{fulL5}) is our main result.
Most notably, we have removed the electric U(1) gauge 
structure by writing
the Lagrangian in terms of the manifestly invariant 
quantities $B_\mu$ and $\vect$.
We have also exposed a manifest duality between the electric 
variable $\vect$ and the magnetic variables $\vecu$ 
and $\vecv$, which becomes particularly
transparent when we specialize to static 
ground state configurations
described by the ensuing Hamiltonian
\be 
H_{YM} \ = \ 
\frac{1}{4} (H_{ij} + M_{ij} + K_{ij} t_3 )^2 
\ + \ \frac{1}{2} (\partial_i \rho)^2 \ + \ 
\rho^2 (\nabla_i \cdot \vect)^2 \ + \ \rho^2 B_i^2 
\ + \ \frac{1}{8} \rho^4 t_3^2 
\la{fulH1}
\ee
\be
+ \ \frac{1}{2} \rho^2|\partial_i \vecm |^2
\ + \ \frac{1}{2} \rho^2 \left( t_+ Q_i^2 \ + \ 
t_- {\bar Q}_i^2 \ + \ t_3 \epsilon_{ijk} m_i
[ H_{jk} + M_{jk} + K_{jk} t_3] \right)   
\ \ + \ ghosts 
\la{fulH2}
\ee
Here $\vecm$ is a three component unit vector that
emerges in the static limit where $\vecm = 
\vecu = \vecv$, and $Q_i = U_i = V_i$.
This follows when we contract the 
full Euclidean rotation group SO(4) = SU(2)$\times$SU(2)
to the spatial rotation group SO(3).
Notice in particular that in the static limit
\[
K_{ij} \ = \ \vecm \cdot \partial_i \vecm 
\times \partial_j \vecm
\]

The result (\ref{fulL4}), (\ref{fulL5}), or its 
Hamiltonian form (\ref{fulH1}), (\ref{fulH2})
is remarkably similar to (\ref{fada}), {\it including}
the potential term. Indeed, for the vector field
$\vect$ we find the following potential 
from (\ref{fulL4}), (\ref{fulL5})
\be
V(\vect) \ = \  \frac{1}{8} \rho^4 t_3^2 \ + \
\frac{1}{2} \rho^2 ( t_- U_\mu V_\mu
\ + \ t_+ U_\mu^* V_\mu^* \ + \ t_3 [ 
H_{\mu\nu} + M_{\mu\nu} +  
K_{\mu\nu} t_3 ] G_{\mu\nu} )
\la{pot1}
\ee
We note that this is an example of the general class 
of potentials that have been considered in the context
of the baby Skyrme model \cite{zakr}.
The same applies to the potential term
for the vector $\vecm$, for which we get from
(\ref{fulH1}), (\ref{fulH2})
\be
V(\vecm) \ = \ m_i \epsilon_{ijk}
[ H_{jk} + M_{jk} + K_{jk} t_3] t_3   
\ee
Notice that even though these potential
terms do seem to break the global SO(3) rotation
invariance of the vector fields $\vect$ and
$\vecm$, we have manifest SO(3)
covariance. 

We shall now continue with somewhat more speculative
comments on the possible phase structures 
of the Yang-Mills theory in the infrared limit, how
a Lagrangian such as (\ref{fada}) could emerge. 
We start by noting that the Lagrangian (\ref{fulL1}) is quite
reminiscent of a Lagrangian to which the 
Coleman-Weinberg-Savvidy \cite{cole},
\cite{savvi} arguments apply. 
Indeed, one can show \cite{lisa} that 
logarithmic corrections at the one-loop level 
lead to a dimensional
transmutation with $\rho$ acquiring a nontrivial ground
state expectation value
\be
< \! \rho^2 \! > \ = \ \Lambda^2 \ \not= \ 0
\la{sbbw}
\ee
Due to the last two terms
in (\ref{fulL4}) this would imply both the U(1) invariant
vector $B_\mu$ and the vector $\vect$
become massive. 

We proceed by considering the properties 
of the Lagrangian (\ref{fulL4}), (\ref{fulL5})
in a naive derivative expansion where we treat 
each of the variables subsequently as a ``slow'' variable
and then study the response of the 
remaining ``fast'' variables in this background.
That is, we envision a Born-Oppenheimer type
approximation to become applicable.

We first take the electric variable $\vect$ to be a
fast variable in the background of the slow magnetic variables.
For this we note that nontrivial average values 
$<\! \vecu \! >$ and $<\! \vecv \! >$ would imply that
the underlying symmetries become broken. Since these
symmetries relate to rotation symmetry in the Euclidean four-space
which can not become broken, it is reasonable to set
\[
<\! \vecu \! > \ = \ <\! \vecv \! > \ = \ 0
\]
The terms linear in $\vecu$ and $\vecv$ then vanish
to the leading order and we conclude that in the
first approximation when we also replace 
$\rho$ by its expectation value (\ref{sbbw}),
the Lagrangian (\ref{fulL4}), (\ref{fulL5}) 
simplifies into
\be
L_{YM} \ \approx \ 
\frac{1}{4} ( H_{\mu\nu} + \ \vect \cdot 
\partial_\mu \vect \times \partial_\nu \vect 
)^2 \ + \ 
\Lambda^2 (\partial_\mu \vect)^2 \ + \ 
\Lambda^2 B_\mu^2 \ + \ \frac{1}{8} \Lambda^4 t_3^2
\ + \ \frac{1}{2} \Lambda^2 (t_+ S_+ \ + \ t_- S_- )
\la{inf1}
\ee
Here we identify the model (\ref{fada}) in interaction 
with a massive vector field. Note that due to the potential term 
the global SO(3) symmetry 
of $\vect$ becomes transformed into a covariance 
{\it w.r.t.} the background. Note also that this potential term
is a combination of the $t_3$ mass term together 
with the analog of an external magnetic field 
coupling to $t_\pm$.
As such, the potential term is present whenever the 
Coleman-Weinberg-Savvidy argument is 
applicable and the background
$\bfe_\mu$ is not identically constant, as
\be
S_+ \ = \ <\partial_\mu \bfe \cdot \partial_\mu \bfe >  
\ee
We note that the result (\ref{inf1}) strongly suggests
that in the infrared limit the electric phase of 
the SU(2) Yang-Mills 
theory describes the dynamics of massive 
knotted solitons \cite{prl}.

The model (\ref{fada}) also emerges from 
a similar Born-Oppenheimer limit for the dual magnetic 
variables, when we consider them as fast variables in the
background of slowly varying electric variables. 
For this we note that a nontrivial 
expectation value in $<\! \vect \! >$
implies that the underlying global symmetry 
becomes broken. But we
expect that
\[
< \! \vect \! > \ = \ 0
\]
From $\vect \cdot \vect = 1$ we then conclude that
\[
< \! \vect_1^2 \! > \ = \ < \! \vect_2^2 \! > \ = \  < \! 
\vect_3^2 \! > \
= \ \frac{1}{3}
\]
When we average the Lagrangian over $\vect$ we find
to the leading order
\be
L_{YM} \ \approx \ 
\frac{1}{4} K_{\mu\nu}^2 \ + \  
\frac{1}{12} H^2_{\mu\nu} \ + \ \Lambda^2 B_\mu^2
\ + \ \frac{1}{2}\Lambda^2 
( |\partial_\mu \vecu |^2 \ + \ |\partial_\mu \vecv |^2)
\ + \ \frac{1}{6} \Lambda^2 K_{\mu\nu} G_{\mu\nu}
\la{app1}
\ee
and when we specify this to static configurations
we find for the Hamiltonian
\be 
H \ = \ \Lambda^2 (\partial_i \vecm)^2 \ + \ 
\frac{1}{2} (\vecm \cdot \partial_i \vecm 
\times \partial_j \vecm)^2 \ + \ \frac{1}{12} H_{ij}^2
\ + \ \Lambda^2 B_i^2 \ + \ \frac{1}{6} \Lambda^2 m_i 
\epsilon_{ijk} <K_{jk}>
\la{app2} 
\ee
Note that the explicit global SO(3) symmetry in $m_i$ 
becomes broken when the background analog of the 
external magnetic field 
$\epsilon_{ijk}K_{jk}$ is nontrivial: Again, we have 
a potential term that removes massless states in $\vecm$
from the spectrum. A comparison with (\ref{inf1}) 
also reveals a manifest duality between
the electric and magnetic variables. In particular,
we find that both sets of variables lead to a description 
that naturally contains massive knotted solitons in 
the spectrum \cite{prl}

\vskip 0.6cm

In conclusion, we have introduced an explicit realization of
the electric and magnetic variables 
in SU(2) Yang-Mills theory and found that both variables
relate to an effective action of the form (\ref{fada}). 
This symmetric appearance of the electric and magnetic
variables both at the level of the field decomposition, 
and at the level of the effective action then suggest a 
duality structure. In particular, our results
support our earlier proposal that the non-perturbative 
spectrum of the Yang-Mills theory describes 
stable knots which are made out of the confining string.

\vskip 0.5cm

We thank A. Alekseev, D. Diakonov, J. Minahan, S. Nasir,
V. Petrov and P. van Baal for discussions.

\end{document}